\documentclass[12pt]{article}
\def\beq{\begin{eqnarray}}
\def\eeq{\end{eqnarray}}
\def\beqs{\begin{eqnarray*}}
\def\eeqs{\end{eqnarray*}}
\def\dl{\delta}
\newcommand{\be}{\begin{equation}}
\newcommand{\ee}{\end{equation}}
\newcommand{\lll}{\langle}
\newcommand{\rrr}{\rangle}

\newcommand{\T}{\mbox{Tr}\> }

\def\centeron#1#2{{\setbox0=\hbox{#1}\setbox1=\hbox{#2}\ifdim
\wd1>\wd0\kern.5\wd1\kern-.5\wd0\fi
\copy0\kern-.5\wd0\kern-.5\wd1\copy1\ifdim\wd0>\wd1
\kern.5\wd0\kern-.5\wd1\fi}}
\def\ltap{\;\centeron{\raise.35ex\hbox{$<$}}{\lower.65ex\hbox{$\sim$}}\;}
\def\gtap{\;\centeron{\raise.35ex\hbox{$>$}}{\lower.65ex\hbox{$\sim$}}\;}
\def\gsim{\mathrel{\gtap}}
\def\lsim{\mathrel{\ltap}}

\begin{document}
\begin{titlepage}
\begin{flushright}
{ITP-UU-02/63}
\end{flushright}

\vskip 1.2cm

\begin{center}

{\LARGE\bf A remark on the short distance potential in
gluodynamics}

\vskip 1.4cm

{\large  V.I. Shevchenko}
\\
\vskip 0.3cm
{\it Institute for Theoretical Physics, Utrecht University,
Leuvenlaan 4 \\ 3584 CE Utrecht, the Netherlands}
\\
\vskip 0.1cm
and \\
\vskip 0.1cm
{\it Institute of Theoretical and Experimental Physics,
B.Cheremushkinskaya 25 \\ 117218 Moscow, Russia } \\
\vskip 0.25cm
e-mail: V.Shevchenko@phys.uu.nl

\vskip 2cm

\begin{abstract}

The structure of leading nonperturbative corrections to the static
Coulomb potential in QCD at small distances is analyzed. We argue
in favor of the correction linearly dependent on distance and
remark that lattice measurements of static potential for charges
in higher representations can distinguish between different
phenomenological models explaining its existence. Related problems
of validity of Dirac quantization condition for running charges in
abelian theory and significance of the quantity $\lll A^2\rrr$ are
briefly discussed.

\end{abstract}
\end{center}

\vskip 1.0 cm

\end{titlepage}

\setcounter{footnote}{0} \setcounter{page}{2}
\setcounter{section}{0} \setcounter{subsection}{0}
\setcounter{subsubsection}{0}


\section{Introduction}

Much attention has been given recently to the question  about
next-to-leading terms for the static potential in confining theory
such as QCD at short distances, see \cite{az,zc,p1,p3,s1,badalian}
and references therein. This problem is closely related to the
structure of OPE in confining theories, despite to define the
potential one has to go to the large time limit and leave
therefore the region of applicability of the standard OPE
\cite{wilson}. In particular, the appearance of new terms in the
so called SVZ condensate expansion \cite{svz} is discussed
\cite{zc}. Phenomenologically, the static potential between charge
and anticharge is given at small distances as the following
expansion \be V(r) = \frac{c_{-1}}{r} + c_0 + c_1 r + c_2 r^2 +
{\cal O} (r^3) \label{ee1} \ee where the coefficients $c_i$ may
also depend on $r$ but only logarithmically. Some typical patterns
look as follows
\begin{itemize}
\item Abelian charges $(e;-e)$ in infinite space beyond tree level
\be c_{-1} = - \frac{e^2(r)}{4\pi} \;\;\; ; \;\;\; c_1 = c_2 =...
= 0 \label{ee11} \ee \item Abelian charges $(e;-e)$ in the cavity
of the size $L$, tree level \cite{cpol} \be c_{-1} = -
\frac{e^2}{4\pi} \;\; ; \;\; c_1 = 0 \;\; ; \;\; c_i = \gamma_i
e^2 L^{-(i+1)} \;\; {\mbox{for}} \;\; i \ge 2 \label{ee12} \ee
\item Nonabelian charges $(gT_D^a;-gT_D^a)$ in the representation
$D$ of $SU(N)$ interacting with soft nonperturbative fields
\cite{ds,sty} \be c_{-1} = - \tilde\alpha_s = - C_D\> (g^2/4\pi)
\;\; ; \;\; c_1 = 0 \;\; ; \;\; c_2 = \gamma \tilde\alpha_s \lll
G_{\mu\nu}^a G_{\mu\nu}^a \rrr T_g \;\; , ... \label{ee13} \ee
\end{itemize}
where in the last case $\lll G_{\mu\nu}^a G_{\mu\nu}^a \rrr$ is
nonperturbative gluon condensate \cite{svz} and nonperturbative
correlation length $T_g$ characterizes the fall-off of the
gauge-invariant two-point correlator of the field strength tensors
normalized to the gluon condensate at the origin (see review
\cite{ourrev} and references therein). The eigenvalue of quadratic
Casimir operator reads as \be C_D = \T_D T_D^a T_D^a \;\;\; ;
\;\;\; \T_D {\bf 1}_D = 1 \label{e2} \ee In the expressions above
$\gamma$'s stay for some dimensionless numerical factors whose
values are of no relevance for our discussion. What attracts the
attention in all mentioned cases is the absence of linear
correction proportional to $c_1$. This is also true for another
models \cite{bal} which we have not included in the list.

However, one cannot propose strong theoretical arguments why $c_1$
should vanish in the context of QCD or another theory with
complicated vacuum structure. Moreover, there are at least two
independent sets of lattice data indicating that $c_1 \neq 0 $
\cite{renzo,bali}. Different models explaining small distance
linear correction have been proposed (see [1-6] and references
therein). They can be divided into three groups. In the first
group there are models \cite{zc,s1,badalian} which are based on
different corrections to the perturbative two-point correlation
function - either in the form of gluon propagator $\lll
A_{\mu}^a(x) A_{\nu}^b(y) \rrr$ or in the form of gauge-invariant
path dependent field strength correlator
 $ \lll \T G_{\mu\nu}(x) \Phi(x,y) G_{\rho\sigma}(y) \Phi(y,x)
\rrr $. The second group is represented by the "dynamical cavity"
model \cite{az}, which essentially explores (\ref{ee12}) but with
$L=L(r)$ instead of a constant. Finally, studies of the static
potential at small distances in abelian confining theories
\cite{p1} form the third group. We briefly comment on the latter
one at the end of the paper, while our attention is focused on the
first and the second groups. The main point of our remark is that
despite many common points, it is possible to make distinction on
the lattice between models of the first and of the second group by
studying the small distance static potential for charges in higher
representations of the gauge group.

The paper consists of two relatively independent parts. In the
first part (sections 2 and 3) we discuss some elements of
short-distance physics in abelian theories. In particular, we
present a short review devoted to the following question: does
Dirac quantization condition for bare charges $eg=2\pi$ stay
intact when loop effects are included? There have been long
discussion of this question in the literature, see \cite{pmon} and
references therein. The affirmative answer is given and physics
behind it is explained. We will also point the reader's attention
in the section 3 that local average of the square of the photon
field $\lll A_{\mu}(x) A_{\mu}(x) \rrr $ can be defined in
covariant gauge in the context of theory with boundaries.

The second part of the paper (sections 4 and 5) is concentrated on
the nonabelian theory. We demonstrate the difference between
"dynamical cavity"
--- like model incorporating Meissner effect, and "modified
propagator" -- like models in the section 4, which makes it
possible to distinguish these scenarios. The section 5 presents
our conclusions.

In the rest of the paper unless explicitly stated otherwise, we
work in Euclidean space with the notation for the four-vectors $k
= (k_1, k_2, k_3, k_4)$ and scalar product $kp =
k_{\mu}p^{\nu}\dl^{\mu}_{\nu}$. The three vectors are denoted as
${\bf k} = (k_1, k_2, k_3)$ and the Wick rotation corresponds to
the replacement $k_4 \to ik_0$.

\section{Elements of short-distance physics in Abelian theory
with monopoles}

Since we are interested in the physics of interactions at small
distances, as a warm-up example, we consider here the theory of
abelian vector field $A_{\mu}$ interacting with the massive
electrically charged matter field and external monopole currents.
An integral part of this section is of mini-review type since most
of the discussed results can be found in the literature. We
believe however that consistent presentation of them can be of
some use for the reader.

 After integrating matter fields out, the partition
function of effective low-energy theory is given by \be Z[j^e_\mu]
= \int {\cal D} A_{\mu} \exp\left(-\frac12 \int d^d p A_{\mu}(p)
[D^{-1}]_{\nu}^{\mu}(p) A^{\nu}(-p) - S_{eff}[A] + i \int d^d p \>
j^e_{\mu}(p) A^{\mu}(-p) \right) \label{ea1} \ee where
$S_{eff}[A]$ contains interaction terms for the field $A_{\mu}$.
The function $[D^{-1}]_{\nu}^{\mu}(p)$ can be decomposed as \be
[D^{-1}]^{\mu}_{\nu}(p) = \dl^{\mu}_{\nu} d_0(p^2) +
p^{\mu}p_{\nu} d_1(p^2) \label{ea2} \ee where the function
$d_1(p)$ is gauge-dependent if the theory in question is a gauge
theory. In the latter case the theory can be rewritten in the
field-strength formulation \cite{halpern} as $$ Z[j^e_{\mu\nu}] =
\int {\cal D} F_{\mu\nu}\> \dl(\partial_{\nu}{\tilde F}_{\mu\nu})
\>\exp \left(-\frac18 \int d^d p F^{\mu\nu}(p)
\Delta_{\mu\nu}^{\rho\sigma}(p) F_{\rho\sigma}(-p) \right)\cdot
$$
\be \cdot\exp\left( - S_{eff}[F] + i \int d^d p \> j^e_{\mu\nu}(p)
F^{\mu\nu}(-p) \right) \label{ea3} \ee where ${\tilde F}_{\mu\nu}
= \frac12 \epsilon_{\mu\nu\alpha\beta} F_{\alpha\beta}$ and \be
\Delta_{\mu\nu}^{\rho\sigma}(p) = (\delta_{\mu}^{\rho}
\dl_{\nu}^{\sigma} - \delta_{\mu}^{\sigma} \dl_{\nu}^{\rho})
\Delta_0(p^2) + (p_{\mu}p^{\rho} \dl_\nu^\sigma -
p_{\mu}p^{\sigma} \dl_\nu^\rho + p_{\nu}p^{\sigma} \dl_\mu^\rho -
p_{\nu}p^{\rho} \dl_\mu^\sigma) \Delta_1(p^2)
 \label{ea4} \ee
The relation \be d_0(p^2) = p^2 (\Delta_0(p^2) + p^2
\Delta_1(p^2)) \label{ea5} \ee provides correlator's matching
between the two formulations.

The field strength formulation (\ref{ea3}) allows one to introduce
monopoles into the theory performing the shift  \be
\dl(\partial^{\nu}{\tilde F}_{\mu\nu}) \to
\dl(\partial^{\nu}{\tilde F}_{\mu\nu} -  j^m_\mu) \label{ea6} \ee
If one may neglects $S_{eff}[F]$ (see discussion below),
straightforward integration gives the resulting monopole partition
function: \be Z[j^m_{\mu}, j^e_{\mu\nu} = 0] = \int {\cal D}
B_{\mu} \exp\left(-\frac12 \int d^d p B_{\mu}(p) \frac{\dl^\mu_\nu
p^2 - p^\mu p_\nu}{\Delta_0(p^2)} B^{\nu}(-p) + i \int d^d p \>
j^m_{\mu}(p) B^{\mu}(-p) \right) \label{ea7} \ee

For the free theory $d_0(p^2) = p^2 \> , \> \Delta_0(p^2) = 1 \> ,
\> \Delta_1(p^2) = 0$, which means that (\ref{ea1}) coincides with
(\ref{ea7}) up to the interchange of electric and magnetic
currents (charges), in other words pure photodynamics is exactly
self-dual. If we switch on the interactions, the inverse
propagator gets renormalized and runs with $p^2$ as $d_0(p^2) =
p^2(1+\Pi(p^2))$ where $\Pi(p^2)$ is the corresponding
polarization operator. Let us assume for the time being that the
dynamics keeps the gauge invariance of the theory intact, which
means, in particular that $\Pi(0)=0$. When computing the static
potential in $d=3+1$ dimensions, one gets formally in the electric
case \be V_{e\bar e} (r) = - e^2 \int \frac{d^{3} {\bf p}
}{(2\pi)^{3}} \> \frac{1}{d_0({\bf p}^2)}\> \exp(i{\bf p}{\bf r})
 \label{ea8} \ee
while corresponding expression in the magnetic case reads \be
V_{m\bar m} (r) = - g^2 \int \frac{d^{3} {\bf p} }{(2\pi)^{3}} \>
\frac{{d_0({\bf p}^2)}}{{\bf p}^4} \>  \exp(i{\bf p}{\bf r})
 \label{ea81} \ee
where the charges $e,g$ are physical renormalized (but not
running) charges and $r=|{\bf r}|$.

It is seen from (\ref{ea8}), (\ref{ea81}) that the leading
corrections to the Coulomb potential have different signs in
electric and magnetic cases: \be \frac{1}{e^{2}}\>\delta V_{e\bar
e}(r) = - \frac{1}{g^{2}}\> \delta V_{m\bar m}(r) = \int
\frac{d^{3} {\bf p} }{(2\pi)^{3}} \> \frac{\Pi({\bf p}^2)}{{{\bf
p}^2}}\> \exp(i{\bf p} {\bf r})
 \label{ea82} \ee

Sometimes this is considered as a proof\footnote{Actually only at
the next-to-leading order in electric coupling $e^2$, which is
assumed to be a small parameter. Notice that $\Pi(p^2)$ is ${\cal
O}(e^2)$.} of the fact that Dirac quantization condition $eg =
2\pi n$ holds also for running charges (see discussion and further
references in \cite{pmon}) if it does for the bare charges: \be
e_0g_0 = eg = e(r)g(r) = 2\pi n \label{ea83} \ee where $e_0 , e,
e(r)$ denote the bare, the renormalized and the running charge,
respectively. The situation is much more subtle, however. To be
more concrete let us take the best studied case of spinor
electrodynamics as an example. The problem is that introducing
monopoles by (\ref{ea6}) one sets the scale of magnetic fields by
the large coupling $g\sim 1/e$, since $j^m_{\mu}$ contains factor
$g \gg 1$. As an immediate result, all higher order irreducible
correlators of magnetic fields interacting with the electrically
charged matter field described by $S_{eff}$ can play a role. One
should distinguish three different kinematical regions. For large
distances $rm \gg 1$, where $m$ is electron mass the corrections
to the static potential are exponentially damped with $r$ for both
$V_{e\bar e}(r)$ and $V_{m\bar m}(r)$. Of main interest is the
small distance region $rm \ll 1$. The leading term in the
$k$-point irreducible vertex in $S_{eff}$ contributing to the
monopole potential (\ref{ea81}) at small distances contains the
factor $g^k e^k = (2\pi n)^k $ while the analogous contribution to
(\ref{ea8}) is suppressed as $e^k e^k = e^{2k} \ll 1$. On the
other hand the $k>4$ correlators describing the processes of
multi-photon scattering do not have\footnote{As is well known, at
$k=4$ the amplitude formally diverges logarithmically, but this
divergence is exactly cancelled in the sum of all diagrams}
contributions logarithmically rising with $p^2$. It means that for
large $p^2$ and not too large value of the product $eg$ one can
indeed take the one-loop result (\ref{ea82}) as a correct first
approximation. With the product $eg$ getting large, however,
higher order terms in the non-Gaussian part $S_{eff}$ are becoming
more and more important and their effect might overcome the
leading kinematical one-loop logarithm and spoil (\ref{ea83}).
However, this is not what happens. Since the effective Lagrangian
is formed in this case at the distances $l\sim r / \sqrt{eg} $
which are for $eg \gg 1$ smaller that the typical scale of the
field change $r$ (see, e.g. \cite{gt,ritus}), the constant field
approximation can be used. The Lagrangian in constant magnetic
field is given (we use Minkowskii metric conventions here) at the
first order by the following expression \cite{ge} \be L = -
\frac{{\bf H}^2}{2}\left( 1 - \frac{\alpha}{3\pi} \log \frac{e
{\mbox{H}}}{m^2} \right)\label{ea13} \ee where $4\pi \alpha = e^2$
and ${\mbox{H}} = |{\bf H}|$. Since the point-like magnetic charge
is defined microscopically via divergence of ${\bf H}$,
${\mbox{div}}\> {\bf H}({\bf r}) = g \cdot \dl ({\bf r})$, one
gets from (\ref{ea13}) for the running magnetic charge with
logarithmic accuracy \be g(r) = g\left[1 - \frac{\alpha}{3\pi}
\log \frac{1}{(mr)^2} \right]\label{ea14} \ee On the other hand,
point-like electric charge is defined via Maxwell's equation
${\mbox{div}}\> {\bf D}({\bf r}) = e \cdot \dl ({\bf r})$ where
${\bf D} =
\partial L /
\partial {\bf E}$, which leads to the standard Uehling-Serber result
\cite{ues} (again with logarithmic accuracy) \be e(r) = e\left[1 +
\frac{\alpha}{3\pi} \log \frac{1}{(mr)^2} \right]\label{ea15} \ee
This is precisely the same answer one can get from (\ref{ea82})
with the standard one-loop expression for $\Pi({\bf p}^2)$.

At first glance the coincidence between (\ref{ea82}) and
(\ref{ea14}), (\ref{ea15}) seems rather surprising since the
latter two expression were obtained from the exact strong field
Lagrangian (\ref{ea13}) which sums up infinite number of graphs
while the former one is a trivial outcome of Gaussian integrations
with one-loop two-point function. The reason for this result is
the remarkable correspondence between QED at small distances and
QED in strong fields, studied in the series of papers
\cite{ritus}. It follows directly from the minimal coupling
principle for the gauge and charged matter fields: $p_{\mu} \to
P_\mu = p_\mu - eA_{\mu}$ which gives a hint that large $p_{\mu}$
and large $eA_\mu$ can describe one and the same physics. The
above example shows rather nontrivial manifestation of this
correspondence. It is worth saying that the transition from
one-loop dominated regime (\ref{ea82}) to truly nonperturbative
one (\ref{ea13}) at fixed distance $r$ with the increasing of
$n=\frac{eg}{2\pi}$ is a physical phenomenon, not related to any
formal redefinitions of coupling constants etc. Notice also that
no topological objects like Dirac strings have been involved in
our analysis. Unfortunately, up to the author's knowledge there
still exists no rigorous proof of (\ref{ea83}) in QED in all
orders of the electric coupling and/or beyond logarithmic
accuracy.

If ${\tilde \Pi}(p^2) \equiv p^2 \Pi(p^2)$ does not vanish when
$p^2 \to 0$, the theory leaves Coulomb phase and one cannot
introduce external monopoles simply by (\ref{ea6}). A well known
example of such theory is given by abelian Higgs model where the
electrically charged field is condensed: \be L= \frac14 F_{\mu\nu}
F^{\mu\nu} + \frac12 |D_{\mu}\phi|^2 + \lambda (|\phi|^2 -
\eta^2)^2 \label{ea32} \ee Attempts to write down exact expression
for the confining potential in this theory encounter a serious
problem. The reason for that is the physical one: since the
confining string is created between the particles (external
monopole-antimonopole pair), its quantum dynamics should be
properly taken into account. The confining interaction can not
therefore be described in terms of particle exchanges. Moreover,
even in effective abelian theory framework there exists no
consistent procedure to perform corresponding stringy calculations
analytically at the moment. The best one can do is to compute the
Wilson loop for a particular choice of the confining string
world-sheet geometry. The conventional choice is the minimal
surface $S$ for the given contour $C$, the flat one if the contour
is rectangular (as it is for the static potential). In the London
limit the Higgs boson is much heavier that the vector boson, $m_H
\gg m=e\eta$ and such defined potential reads in this limit (see,
e.g. \cite{ach}): \be V^{[S]}_{m{\bar m}}(r) = g^2 \lim_{T\to
\infty} \frac{1}{T} \> \int d^4 x \int d^4 y \left(\frac{1}{4} \>
\Sigma_{\mu\nu}(x) \Delta_m(x-y) \Sigma_{\mu\nu}(y)\right) - g^2
\int \frac{d^3 {\bf p}}{(2\pi)^3} \frac{\exp(i{\bf p}{\bf
r})}{{\bf p}^2 + m^2} \label{ea20} \ee where the surface current
$\Sigma_{\mu\nu}(x)$ and the kernel $\Delta_m(x-y)$ are given by
the expressions \be \Sigma_{\mu\nu}(x) = \int\limits_S
d\sigma_{\mu\nu}(\xi) \> \dl^{(4)}(x(\xi) - x) \;\; ; \;\;
\Delta_m (x) = \int \frac{d^4 p}{(2\pi)^4} \> \frac{m^2}{p^2 +
m^2} \> \exp(ipx) \label{ea21} \ee We have indicated by
superscript$^{[S]}$ that the potential depends on the profile of
the string chosen in (\ref{ea21}). At large distances $rm \gg 1$
the potential (\ref{ea20}) has linear asymptotics and describes
confinement.

We are to address the question of applicability of (\ref{ea20}) at
small distances. Obviously, the expression (\ref{ea20}) breaks
down at $r \lsim m_H^{-1}$ since at such small distances the
scalar field condensate gets excited and virtual loops of scalar
particles contribute to the polarization of the vacuum. If one
takes (\ref{ea32}) as some low energy effective theory, the
ultraviolet cut-off implicitly enters through higher-dimensional
operators. Moreover, if the Higgs boson in this effective theory
is composite in terms of underlying microscopic theory degrees of
freedom, the vector bosons with the energy higher than the
corresponding binding energy start to resolve the constituents. In
any of these cases the potential (\ref{ea20}) gets modified. Let
us consider the simplest possible modification which is to
replaces $m^2$ in (\ref{ea20}), (\ref{ea21}) into vector boson
self-energy ${\tilde \Pi}(p^2)$ such that ${\tilde \Pi}(0) = m^2$.
Such replacements is in line with (\ref{ea8}), (\ref{ea81}) and
can be justified\footnote{Of course, ${\tilde\Pi}(0) = 0$ in the
perturbative QED context} in terms of perturbation theory in the
electric coupling $e$. Needless to say that the high-momentum
asymptotics of ${\tilde \Pi}(p^2)$ will be different for different
microscopic scenarios outlined above.

The leading short-distance contribution to (\ref{ea20}) is
attractive Coulomb potential while there are two kinds of
sub-leading corrections: \be V^{[r\times T]}_{m\bar m} (r) =
-\frac{g^2}{4\pi r} + \dl V^{(1)}_{m\bar m}(r) + \dl
V^{(2)}_{m\bar m}(r) + {\cal O}(g^2 e^4) \label{ea35} \ee
 The second correction term comes from the second
Yukawa-like term in (\ref{ea20}) and formally reads as \be
\frac{1}{g^2} \dl V^{(2)}_{m\bar m}(r) = \int \frac{d^3 {\bf
p}}{(2\pi)^3} \frac{{\tilde\Pi}({\bf p}^2)}{{\bf p}^4}\>
\exp(i{\bf p}{\bf r}) \label{ea34} \ee The low - ${\bf p}^2$
divergence is artificial and can be easily regularized. The first
term in (\ref{ea20}) produces the following correction \be
\frac{1}{g^2} \dl V^{(1)}_{m\bar m}(r) = r \cdot \int \frac{d^3
{\bf k}}{(2\pi)^3} \frac{{\tilde\Pi}\left(\frac{{\bf
k}^2}{r^2}\right)}{{\bf k}^2}\> \left(
\frac{4\sin^2\left(\frac{k_3}{2}\right)}{(k_3)^2}\right)
\label{ea342} \ee where for convenience we put the Wilson contour
in the $(3,4)$ plane. It is seen that the dependence of the
confining term - induced correction (\ref{ea342}) on $r$ is
completely determined by the behavior of $\tilde\Pi(p^2)$ as a
function of $p^2$. In particular, if $\tilde\Pi(p^2)$ is such that
the $r$ - dependence of the integral is weak enough, the
correction to the potential will be approximately linear. If
$\tilde\Pi(p^2) \equiv m^2$, the Yukawa-like correction is linear
in $r$ and negative: $\dl V^{(2)}_{m\bar m}(r) = - (g^2 m^2
r)/(8\pi)$ (we have omitted irrelevant constant term). This is the
only correction (up to the change $g\to e$) one would get for the
static electric charges without taking into account the string
dynamics. The "confining" correction (\ref{ea342}) is positive in
this case, but logarithmically divergent in the ultraviolet
region. This divergence is to be cured if the correct form of
${\tilde \Pi}(p^2)$ in high-momentum regime is taken. Presumably,
the relative minus sign between $\dl V^{(1)}_{m\bar m}(r)$ and
$\dl V^{(2)}_{m\bar m}(r)$  persists in this case as well.

It is worth to put an emphasis on the fact that the physics behind
$\dl V^{(1)}_{m\bar m}(r)$ and $\dl V^{(2)}_{m\bar m}(r)$ is very
different. In a sense, the former one describes the process of
string tension formation, which starts essentially at some
ultraviolet scale in this model and goes all the way up to the
distance scale given by $m^{-1}$. Since the corresponding
formation "speed" is only logarithmic, one can have indeed
approximately linear potential at small distances - "ultraviolet
confinement", the phenomenon which apparently goes beyond the
conventional OPE.

\section{Comment on the quantity $\lll A^2 \rrr$ }

The linear correction to the potential imply the existence of
local $d=2$ parameter in the theory proportional to $c_1$. It
cannot be local polynomial function of original fields of the
theory since the only dimension two candidate, average of vector
potential squared $\lll \T A_\mu(x)^2 \rrr$ is not
gauge-invariant. Moreover, there is a gauge, defined by
infinitesimal transformation $U(x) = 1 +
igA_{\mu}(x_0)(x-x_0)^{\mu} +..$, such that $^U\!A_{\mu}^a(x_0) =
0$, i.e. one always has locally \be \left\lll \> \min\limits_{U\in
G} \> \left(^U\!A_\mu^a(x)\right)^2 \right\rrr = 0\ee where $G$ is
the gauge group, for abelian or nonabelian theory. In abelian
case, however, the minimum of integral of $\lll A_\mu(x)^2 \rrr$
over the whole space can be rewritten in terms of gauge-invariant
quantities \cite{zzz}. The physical relevance of such nonlocal
object for nonabelian case is under debate \cite{l,zvb}.

We would like to point the reader's attention to the fact, that
despite it seems to be impossible to assign any gauge-invariant
meaning to the local quantity $\lll A^2 \rrr$, one can have in
some cases $\xi$-independent definition of $\lll A^2 \rrr$, where
$\xi$ is the covariant gauge fixing parameter. It happens when one
studies the theories which contain some external parameter of the
dimension of mass/length and if the part of the effective action
depending on this parameter does not depend on $\xi$.
 By way of example let us study
photodynamics with the Casimir-type static boundary conditions
(see review \cite{bordag} and references therein): \be Z[j^e_\mu]
= \int {\cal D} A_{\mu} \dl[BC] \exp\left(-\int d^d x \left(
\frac14 F_{\mu\nu}(x)F^{\mu\nu}(x) +
\frac{1}{2\xi}\left(\partial_{\mu}A_{\nu}(x)\right)^2 + i
j^e_{\mu}(x) A^{\mu}(x) \right)\right) \label{ea22} \ee where the
boundary conditions read as
$$\dl[BC] = \dl(n_{\mu}{\tilde F}_{\mu\nu}(x_1,x_2,x_3 =
a_1,x_4))\dl(n_{\mu}{\tilde F}_{\mu\nu}(x_1,x_2,x_3 = a_2,x_4))
$$
Here
 $n_{\mu} = (0,0,1,0)$ is a unit vector in the
$x_3$-direction and $a_1$ , $a_2$ mark the $x_3$ - positions of
parallel infinitely thin ideally conducting plates. The distance
between plates is given by $a=|a_2 - a_1|$.

The gauge field propagator can be readily obtained from
(\ref{ea22}), it consists of two parts:
$$
\lll A_{\mu}(x) A_{\nu}(y) \rrr = D_{\mu\nu}(x,y) =
D^{(0)}_{\mu\nu}(x-y; \xi) + {\bar D}_{\mu\nu}(x,y; a_1,a_2)
$$
where the term $D^{(0)}_{\mu\nu}(x-y; \xi)$ is the standard
gauge-dependent tree-level photon propagator, while the term
${\bar D}_{\mu\nu}(x,y; a_1,a_2)$ encodes the information about
the boundaries. The exact form of the latter was first obtained in
\cite{bordag2}. The function $D^{(0)}_{\mu\nu}(x-y; \xi)$ is
translation-invariant but gauge-dependent, while the function
${\bar D}_{\mu\nu}(x,y; a_1,a_2)$ depends on $x_3$ and $y_3$
separately, but is $\xi$-independent (not to be confused with the
gauge-invariance!). Moreover, it has finite limit when $x$
approaches $y$. Therefore one can address the issue of the $\lll
A_{\mu}(x)A^{\mu}(x) \rrr$ condensate in the Casimir vacuum
exactly in the same way one computes the energy density in this
problem, namely, subtracting the boundary-independent part: \be
\lll A^2(x) \rrr \stackrel{def}{=} \lim\limits_{y\to x} \left(\lll
A_{\mu}(x)A^{\mu}(y) \rrr - \lll A_{\mu}(x)A^{\mu}(y) \rrr^{(0)}
\right) = {\bar D}_{\mu}^{\mu}(x,x;a_1,a_2) \label{ea25} \ee It is
convenient to introduce the notation $z=a_1 + a_2 - 2x_3$. When,
using exact expression for ${\bar D}_{\mu\nu}(x,y; a_1,a_2)$ one
gets: \be \lll A^2(z) \rrr = \frac{1}{12 a^2}\left[ 1 +
\frac{3}{2}\tan^2\left(\frac{\pi z}{2a}\right) \right]
\label{eq25} \ee Such defined local dimension two "condensate" is
$\xi$-independent, strictly positive and diverges on the
boundaries. In completely analogous way one can define $\lll A^2
\rrr$--"condensate" at finite temperature.

In the context of Casimir problem one is usually interested to
study the changes of physical quantities like components of
energy-momentum tensor with respect to their vacuum values. The
above example demonstrates that "non-physical" quantity like $\lll
A^2 \rrr$ is also nontrivially modified. It is worth noting that
(\ref{eq25}) is not related to any linear correction to the
potential - the potential for dipole between the mirrors is of the
form (\ref{ee12}) with $L=a$.

\section{"Dynamical cavity" vs "modified propagator"}

One of the main assumptions of the SVZ framework can be formulated
as a wordplay: the nonperturbative vacuum of the theory is not
perturbed by the external sources. In case of large virtualities
it is sometimes justified using the language of condensed matter
physics: since the time scale of the hard process we wish to study
is much smaller than typical relaxation times characterizing such
a medium as nonperturbative QCD vacuum, it is reasonable to
suggest that the vacuum state remains unchanged.

It is of interest to relax this assumption and to study the
corresponding physics. An example of the situation where one has
to do it is provided by the monopole-antimonopole pair in the
superconductor. It is well known that the condensate must be
broken along some line connecting the particles, whatever small
their charge is (see review \cite{pmon} and references therein).
The reason for that, eventually, is the nonperturbative nature of
the interaction between magnetic and electric particles.

First, we re-derive the results of "dynamical cavity" model
\cite{az} in more quantitative way. After that, we will come to
the "modified propagator" model \cite{zc}.  Since we consider
gluodynamics in this section, we switch to dual terminology, so
that particles which are confined (quarks) carry {\it electric}
and not magnetic charge. The microscopic description of
confinement as dual Meissner effect via monopole condensation
refers to the abelian projection procedure \cite{tho}. We adopt
the physical picture of confinement as dual Meissner effect but
will make no use of the abelian Higgs model--motivated language in
this section.
 As it is well known the energy density of
nonperturbative fields is given by the energy-momentum tensor
trace anomaly \cite{qq} \be \epsilon = \frac14 \lll
\theta_{\mu\mu} \rrr = \frac{\beta(\alpha_s)}{16 \alpha_s} \lll
G_{\mu\nu}^a(0)G_{\mu\nu}^a(0) \rrr \label{ea96} \ee where \be
\beta(\alpha_s) = \mu^2 \frac{d\alpha_s(\mu^2)}{d\mu^2} =
-\left(\frac{11}{3}N - \frac{2}{3}N_f\right)
\frac{\alpha_s^2}{2\pi} + {\cal O}(\alpha_s^3) \label{ea97} \ee

In principle the average (\ref{ea96}) depends on the actual state
of the system, for example, one can study its density or
temperature dependence. In terms of invariant functions $D(z^2)$
and $D_1(z^2)$ the vacuum gluon condensate is given as \cite{ds}
\be  \lll g^2\T G_{\mu\nu}(0) G_{\mu\nu}(0) \rrr = 24(D(0) +
D_1(0)) \label{ea94} \ee where only the function $D(z^2)$ is
responsible for confinement. In particular, the deconfinement
transition is characterized by the condition $D(z^2) \equiv 0$,
with the smooth behavior of $D_1(z)$ over the phase transition
\cite{dig}.

One can rise a question what happens to (\ref{ea96}) in color
field of external source. Owing to $SU(3)$ and $O(4)$ invariance,
such average can be defined as \be G_2({\bf r}) =  \left\lll
\frac{\alpha_s}{\pi} G_{\mu\nu}^a({\bf r})G_{\mu\nu}^a({\bf r})
\right\rrr =  \frac{g^2}{2\pi^2} \> \frac{\left\lll \T (
G_{\mu\nu}({\bf r})  G_{\mu\nu}({\bf r}))\cdot \T {\mbox{P}}\exp
i\int_C A_{\rho}dz_{\rho} \right\rrr}{\left\lll \T {\mbox{P}}\exp
i\int_C A_{\rho}dz_{\rho}\right\rrr} \label{ee16} \ee where $C$
stays for the (closed) world-line of the source. If the point
${\bf r}$ is far enough from $C$, the average reaches the vacuum
value: $G_2({\bf r}) \to G_2$. Correspondingly, one defines
$\epsilon({\bf r}) = \epsilon({\bf E}({\bf r}))$ via (\ref{ea96}).
 To the best of author's knowledge, no systematic lattice
analysis of behavior of $\epsilon({\bf r})$ in the strong external
field has been performed. This problem is of direct relevance for
our discussion.  Namely, let us put a single static charge in the
nonperturbative QCD vacuum. Due to confinement property there
should be an anti-charge somewhere in the space to end the
confining string and therefore the field distribution will not be
spherically symmetric. Suppose, however, that we look at the gluon
fields in a thin spherical layer between $r$ and $r+\Delta r$,
where $r$ is smaller that the typical nonperturbative nonlocality
scale given by $T_g$ (it is worth mentioning that the same
quantity $T_g$ defines the width of the confining string). Then it
is natural to assume that the Faraday flux lines of the charge are
affected only weakly by the opposite charge sitting on the other
end of the string. On the other hand, they are affected by the
nonperturbative vacuum and act back on it. Correspondingly, we
take the energy of the gluon fields as a sum of perturbative and
nonperturbative parts \be {\cal E} = {\cal E}_{p} + {\cal E}_{np}
= \frac12 \> \int\limits_{V} d^3 {\bf r} \> {\bf E}^a({\bf r}){\bf
E}^a({\bf r}) + \int\limits_{V} d^3 {\bf r} \> {\epsilon}({\bf r})
\label{ea95} \ee where ${\bf E}^a({\bf r})$ is the perturbative
electric field of the charge. In principle, the coordinate
dependence of ${\bf E}^a({\bf r})$ and $\epsilon({\bf r}) =
\epsilon({\bf E}({\bf r}))$ on ${\bf r}$ can be very complex,
resembling the formation of intermediate state in the vicinity of
the charge, like it happens in ordinary superconductors. It can be
energetically advantageous to squeeze slightly perturbative flux
lines at one part of the volume\footnote{Since we are working at
small distances, the total perturbative flux (i.e. the number of
the flux lines) is conserved. More accurately, because of
asymptotic freedom it is not, but its violation due to asymptotic
freedom is weak logarithmic effect} and to rarefy them in the rest
of it (this increases the first term in (\ref{ea95})) but
compensate the excess of energy by the nonperturbative energy gain
from the rest of the volume, which can be possible if
$\epsilon({\bf r})$ decreases towards its vacuum value when ${\bf
E}^a({\bf r})$ decreases in this part of the volume. This is the
essence of Meissner effect. Let us simplify the matter further by
approximating (\ref{ea95}) as \be \Delta {\cal E} = \frac12 \>
\left[C_D \> \left(\frac{g}{4\pi r^2}\right)^2 \right]\cdot
\left(\frac{4\pi r^2}{S_1(r)}\right)^2 \cdot S_1(r)\Delta r +
\epsilon_1 \cdot S_1(r)\Delta r + \epsilon_2 \cdot S_2(r)\Delta r
+ {\cal O}(\Delta r^2) \label{ea100} \ee We have taken
$\epsilon({\bf r})$ as constants $\epsilon_{1}$ and $\epsilon_2$
in the parts of the volume $S_1\Delta r$ and $S_2 \Delta r$,
respectively. Notice that $S_1(r) + S_2(r) = 4\pi r^2$.
 The expression (\ref{ea100}) assumes that
the parts of the space occupied by the nonperturbative fields with
the (vacuum) energy density $\epsilon_2$ completely expel the
perturbative field flux lines into the regions where the energy
density of nonperturbative fields is given by $\epsilon_1$. The
rescaling of electric perturbative field of the point charge
reflect the flux conservation.

 Since
$S_1(r)$ can be never greater than $4\pi r^2$, there exists
critical radius (corresponding to the stationary point of
(\ref{ea100})) \be r_c^4 = \frac{{\tilde\alpha}_s}{8\pi(\epsilon_1
- \epsilon_2)} \label{ea102} \ee When $r$ reaches $r_c$ from
above, the region of unperturbed vacuum $S_2$ shrinks to zero.
Physically the model describes the formation of a bag of the
radius $r_c$, there perturbative fields are strong enough to
change the nonperturbative vacuum state significantly. We are not
addressing an interesting question at the moment, is there local
deconfinement transition (in any sense of the word) in this volume
(what is natural to think of being based on abelian Meissner
effect analogy). Let us estimate $r_c$ numerically. We use
different sets of data \cite{data} and take for the gluon
condensate \be G_2 = \left\lll \frac{\alpha_s}{\pi} G_{\mu\nu}^a
G_{\mu\nu}^a \right\rrr = (0.014 \div 0.026) {\mbox{GeV}}^4
\label{ea103} \ee Hence for the nonperturbative vacuum energy in
one loop choosing $N=3$, $N_f=2$ we obtain \be \epsilon_1 -
\epsilon_2 = \kappa \epsilon - \epsilon = \left(1 - \kappa \right)
\cdot(4 \div 8) \cdot 10^{-3} \> {\mbox{GeV}}^4 \label{ea108} \ee
The value of the positive factor $\kappa = \epsilon_1 / \epsilon_2
<1$ is unknown and enters as free parameter. Taking
${\tilde\alpha}_s = 0.47$ (corresponding to the charges in
fundamental representation and $\alpha_s(M_\tau)$), we finally get
\be r_c = (1-\kappa)^{-\frac14} \cdot (0.2 \div 0.3)\> {\mbox{Fm}}
\label{ea106} \ee  Unless $\kappa$ is unnaturally close to unity
this rather small value of $r_c$ is compatible with the discussed
picture - had it been much larger than $T_g$, the analysis would
be meaningless.

Let us, following \cite{az}, consider now a small color dipole of
the size $R$. The expression for the energy (\ref{ea95}) stays
intact, but the physics of critical distance $r_c$ is different.
Since there is no flux of perturbative field through the surface
of the volume with the dipole inside, it is advantageous to
confine flux lines inside some finite volume. Suppose that we have
chosen the volume of some fixed size $L \gsim R$ around the
dipole, such that vacuum energy density $\epsilon_2 = \epsilon$
outside is reduced to $\epsilon_1
> \epsilon_2$ inside it. The perturbative electric field felt on
the boundary of this volume is proportional to $R/L^3$. It is
important that it goes to zero when $R$ decreases. It means that
making $R$ small enough the total energy can also be lowered by
appropriate decreasing of $L$, i.e. $L = L(R)$. Indeed, when we go
from $L$ to $L-\Delta L$ the perturbative energy increase due to
the rearrangement of perturbative field is given by \be \Delta
{\cal E}_p = \frac{1}{2} C_D \left(\frac{g}{4\pi}\right)^2
\int\limits_{L-\Delta L}^{L} d^3 {\bf r} \left( \frac{{\bf
R}^2}{{\bf r}^6} + 3 \frac{({\bf R}{\bf r})^2}{{\bf r}^8} \right)
= {\tilde\alpha}_s \frac{R^2}{L^4} \Delta L  + {\cal O}(\Delta
L^2) \label{ea111} \ee while the nonperturbative energy gain is
\be \Delta {\cal E}_{np} = 4\pi L^2(\epsilon_1 - \epsilon_2)\Delta
L + {\cal O}(\Delta L^2) \label{ea112} \ee This defines the
stationary point \be L_c = \left(\frac{{\tilde\alpha}_s
R^2}{4\pi(\epsilon_1 - \epsilon_2)}\right)^{1/6}\label{ea113} \ee
We can now estimate the nonperturbative correction to the energy
of the dipole as \be \delta {\cal E} (R) \approx \tilde\alpha_s \>
\frac{R^2}{3L_c^3} + (\epsilon_1 - \epsilon_2)\cdot
\frac{4\pi}{3}L_c^3 = \frac{4}{3}
\cdot\sqrt{\pi{\tilde\alpha}_s(\epsilon_1 - \epsilon_2)} \cdot R
\label{ea120} \ee This correction is linear in $R$, as one can
alternatively see inserting (\ref{ea113}) into (\ref{ee12}).
Numerically we get \be \frac{4}{3}
\cdot\sqrt{\pi{\tilde\alpha}_s(\epsilon_1 - \epsilon_2)} =
(1-\kappa)^{\frac12} \cdot (0.10 \div 0.14) \> {\mbox{GeV}}^2
\label{ea780} \ee

The discussed picture reminds the old MIT bag model \cite{jt}.
This similarity is formal, in some sense. While in the MIT bag
model the bag has the typical hadron size, the "dynamical cavity"
of the size $L_c(R)$ given by (\ref{ea113}) is an object, which
has physical meaning only at the very small distances. Since $L_c
\sim R^{1/3}$, at small enough $R$ we are always in $T_g \gsim L_c
\gsim R$ regime, where all the picture is meaningful. With $R$ and
$L_c$ rising and reaching $T_g$, the confining string formation
process starts and we leave the domain of qualitative
applicability of the model.

We can now come to the "modified propagator" models \cite{zc,s1,
badalian}. Actually, their analysis is simpler and we take as an
example the model of \cite{zc}.  It is suggested to modify the
one-gluon exchange propagator by adding "tachyon" mass term to the
gluon, i.e. (in Minkowskii metric and Feynman gauge) \be \lll
A_{\mu}^a(k) A_{\nu}^b(-k) \rrr = {\eta_{\mu\nu}} {\dl^{ab}}
\frac{1}{k^2} \to {\eta_{\mu\nu}} {\dl^{ab}} \left(\frac{1}{k^2} +
\frac{\lambda^2}{k^4} \right) \label{ea67} \ee Notice that
(\ref{ea67}) is large momentum expansion so one actually never
gets closer to the artificial "tachyon pole". The analysis of
existing phenomenology leads to $\lambda^2 \approx -0.5\>
{\mbox{GeV}}^2$ estimate for the "tachyon mass" \cite{zc}. However
the physical origin of the $\lambda^2$--term in the context of QCD
has not been clarified. Alternative scenarios suggest $\mu^2/z^2$
- contribution\footnote{It can be shown that $1/z^2$ term in the
function $D(z^2)$ would produce ultraviolet divergence on the
world-sheet of the confining string, while such term in $D_1(z^2)$
is safe.} to the function $D_1(z^2)$ \cite{s1}, strong coupling
constant freezing \cite{badalian}, infrared renormalons \cite{az}.
It is easy to see that all these proposals lead to the Casimir
scaling law for the corresponding correction, for example, with
expression (\ref{ea67}) one gets \be \ee \be V_D(r) +
{\mbox{const}} = - {\tilde{\alpha}}_s \cdot \left( \frac{1}{r} +
\frac{\lambda^2 r}{2} \right) +  {\cal O}(r^2) \label{ea1077} \ee
where ${\tilde{\alpha}}_s = C_D \alpha_s = C_D (g^2/4\pi)$, i.e.
$V_D(r) \sim C_D$.

The magnitude of corrections are close to each other in both
models, for example taking value (\ref{ea780}) one gets in terms
of (\ref{ea67}) $\lambda^2 = - (1-\kappa)^{\frac12} \cdot (0.4
\div 0.6)\> {\mbox{GeV}}^2$ which is surprisingly close to the
value found in \cite{zc}. The important difference, however is
different $C_D$-dependence. The remarkable property of
(\ref{ea120}) is its square root dependence on the eigenvalue of
Casimir operator $C_D$. This result is known in the context of MIT
bag model \cite{jt2}, where it takes place for the slope of
confining linear potential. As such, square root law was
definitely ruled out by precise recent studies of static potential
at large distances in different representations of the gauge group
\cite{bali5,deldar}. Instead, the Casimir scaling phenomenon
(proportionality of $V_D(r)$ to $C_D$) was confirmed (see
\cite{ourcs} for review). However there exist no calculations of
static potential for charges in higher representations at small
distances . Such lattice simulation will confront in nontrivial
way (\ref{ea120}) and (\ref{ea1077}).

From general point of view the Casimir scaling and $\sqrt{C_D}$
law for the leading small distance correction would correspond to
rather different physical pictures. Both cases physically describe
perturbative-nonperturbative interference and require modification
of the standard "condensate ideology" in QCD. However in the
latter case this modification is to be by far more radical.
Roughly speaking, the former case corresponds to a new kind of
nonperturbative corrections to the perturbative propagator, which
are "harder" than those studied before, while the latter case
indicates that the nonperturbative vacuum structure itself is
strongly affected by the perturbative field when it reaches some
critical value.

To be self-contained, let us briefly mention the analysis of
classical abelian Higgs model on the shortest distances, performed
in \cite{p1}. The small distance correction to the potential was
found numerically and fitted by linear function with a good
accuracy. This behavior was linked to the Dirac veto, i.e. the
condition for the charged condensate $\phi$ to vanish along some
line (the Dirac string) between the dual charges. The problem
mentioned in the section 2 still persists in this case, because
the position of the Dirac string can be chosen in arbitrary way
and the answer for the short distance potential depends on this
choice. Moreover, it is not quite clear what the Dirac string of
effective abelian theory does correspond to in terms of the
original nonabelian theory. In "dynamical cavity" model the linear
behavior (\ref{ea120}) also results from the Meissner effect but
without any special role of singularities like Dirac strings.
Roughly speaking, the effect in \cite{p1} comes from the
infinitely thin line between the charges, while in the discussed
case all the volume of the size $L_c$ contributes to it.

\section{Conclusion}

There are a few models in the literature aimed to reproduce the
linear dependence with distance of the leading nonperturbative
correction to the static potential at small distances. Comparing
different models we call the reader's attention to the fact, that
possible way to put them on test is to check the dependence of
this correction on the quadratic Casimir for higher
representations of the charges. Such computation can shed new
light on strong interaction physics and the problem of
confinement.

{\bf \large Acknowledgments }

\bigskip
 The author is thankful to the foundation
"Fundamenteel Onderzoek der Materie" (FOM), which is financially
 supported by the Dutch National Science Foundation (NWO).
 The support from the grants
RFFI-00-02-17836, RFFI-01-02-06284 and from the grant INTAS 00-110
is also acknowledged.

\bigskip


\end{document}